\documentclass[a4paper]{article}
\usepackage{spconf,amsmath,graphicx,url,psfrag}
 
\title{An Experimental Investigation of SIMO, MIMO, Interference-Alignment (IA) and Coordinated Multi-Point (CoMP)}
\name{Per Zetterberg and Nima N. Moghadam
\thanks{This work was performed partly in the framework of the VINNOVA
sponsored RAMCOORAN project and in the EU-FET project HIATUS}}
\address{ACCESS Linnaeus Center,
KTH Royal Institute of Technology, Osquldas v{\"{a}}g 10, \\
SE-100 44 Stockholm, Sweden, perz@ee.kth.se}
\begin{document}
%
\maketitle
\begin{abstract}
In this paper we present experimental implementations
of interference alignment (IA) and coordinated multi-point transmission
(CoMP). We provide results for a system with three base-stations and 
three mobile-stations all having two antennas. 
We further employ OFDM modulation, with high-order constellations,
and measure many positions both line-of-sight and non-line-of-sight
under interference limited conditions.
We find the CoMP system to perform better than IA
at the cost of a higher back-haul capacity requirement.
During the measurements we also logged the channel estimates
for off-line processing. We use these channel estimates 
to calculate the performance under ideal conditions. 
The performance estimates obtained this
way is substantially higher than what is actually observed in the 
end-to-end transmissions---in particular in the CoMP case 
where the theoretical performance is 
very high.
We find the reason for this discrepancy to be the impact of 
dirty-RF effects such as phase-noise and non-linearities.
We are able to model the dirty-RF effects to some extent.
These models can be used to simulate more complex systems and still
account for the dirty-RF effects (e.g., systems with tens of mobiles
and base-stations). Both IA and CoMP perform better than reference 
implementations of single-user SIMO and MIMO in our measurements.

\end{abstract}
\section{Introduction}
\label{sec:intro}

Interference alignment (IA) is a concept that was introduced in the
seminal paper \cite{CAD:08}. The wording ``interference'' and ``alignment'' 
refers to the fact that
according to the strategy, the interfering signals should be confined
to a subspace disjoint from the subspace of the desired signals, -
when transmitting over MIMO channels. This MIMO channel
may result from using multiple-antennas or using so-called symbol
extended channels (e.g., using multiple carriers). However, in 
this paper we will 
only consider the case of MIMO channels achieved using multiple antennas.

In order to investigate the usefulness of IA in practice, we are herein
analyzing the results of a real-world implementation of IA.
Experimentation with IA was pioneered in \cite{GOL:09,AYA:10} and \cite{GAR:11}.
In the paper \cite{GOL:09} considers a variant of IA with co-operation also
among the receivers. This is different from the scenario herein and will
not be discussed further. The paper \cite{AYA:09} simulations
using measured channels is performed. 
In this paper we are concerned with the difference
between the performance calculated in this manner and the results
that are observed when actually transmitting over the channel.
The paper \cite{GAR:11} does transmit over actual channels. Moreover,
\cite{GAR:11} also presents end-to-end error-vector measurements (EVM) 
which is a very relevant representation of the quality of the end-to-end 
channel, including
the impact of real-world hardware such as non-linearities, phase-noise,
and inter-symbol interference. 
The EVM quantifies the quality of the transmission channel
seen from the view-point of the modulator and de-modulator
which operate over the virtual SISO channels created by
transmit beamformers and receive combiners.

This paper uses an OFDM modulation with higher bandwidth and higher
modulation order than \cite{GAR:11} does. We further include
a high-performance LDPC code and reduce the time between frames
from five seconds used in \cite{GAR:11} to a tenth of a second. 
In addition, we also consider several positions of the mobile-stations
including many non line-of-sight positions thus exposing IA to a more diverse
set of channels.

Parallel to the development of IA, the concept of coordinated multi-point 
transmission (CoMP) has also emerged with interest from 3GPP standardization.
This approach is similar to IA 
but assumes that the signals from multiple base-stations have
a common phase-reference and that all base-stations know the information
to be transmitted to every mobile-station which is not the case in IA. 
However, both IA and CoMP require information about the channel
between all transmitting base-stations and all receiving mobile-stations.

Pioneering experimentation with CoMP is found in 
\cite{JUN:10,DON:11,HOL:11}. 
The paper \cite{JUN:10} presented the performance from trials
using a setup with two LTE base- and mobile-stations.
Solutions for acquiring the necessary channel state information
at the base- and mobile-stations are described.
The paper \cite{DON:11} presents similar measurements. 
While the papers \cite{JUN:10} and \cite{DON:11} present absolute
(impressive) numbers of throughput the papers lack any specific comparison
between the measurements results and theory.
The paper \cite{HOL:11} looks into the difference between
``estimated'' and ``measured'' SINR, but provides very little detail on
their analysis.

In this paper we also implement a form of CoMP. We compare it with IA and 
with the well-known base-line schemes
of single-user SIMO and MIMO. 
We further use three base-stations and mobile-stations
while the above cited papers use two.

Most importantly, we provided a detailed analysis of the difference 
between performance that would have been obtained when performing 
a typical simulation
of the scenario at hand and the performance actually obtained.
This difference (``the delta'') can be used to extrapolate the real-world
performance in more complex scenarios and provide useful insight into
the factors that come into play in the real world.

\section{Testbed Setup}

\begin{figure}
\begin{psfrags}
\psfrag{u2}[b][b][0.5]{signal generator}
\psfrag{a}[b][b][0.5]{antennas}
\psfrag{u}[b][b][0.5]{USRP}
\psfrag{r}[b][b][0.5]{Receiver modules}
\psfrag{d}[t][b][0.5]{Local oscillator generator}
\psfrag{v}[b][t][0.5]{DC generator}
\psfrag{s}[l][bl][0.5]{Cables (10MHz, PPS, Ethernet, power)}
\centerline{
   \includegraphics[width=0.4\textwidth, height=0.25\textheight ]%
     {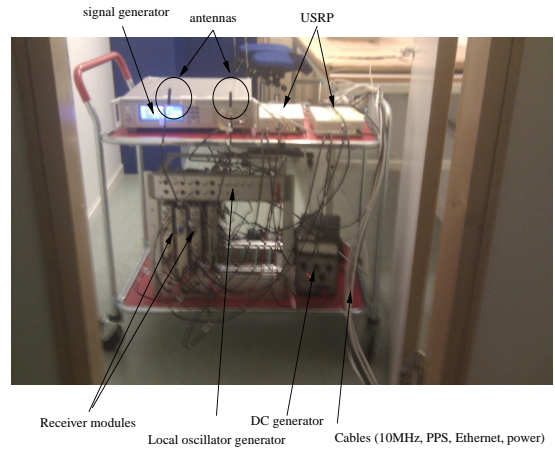}}
\end{psfrags}
\caption{A receiver node }
\label{setup1}
\end{figure}

Our testbed consists of six nodes, of which three take the role
of base-stations and three take the role of mobile-stations.
We consider the downlink but one could also interpret the
results as uplink although it is less natural.
All nodes have two vertically polarized dipole antennas spaced
20cm apart or 1.6 wavelengths at our 2490MHz carrier frequency.
This carrier frequency is unoccupied in the building where the
measurements were conducted.
The base-station transmitters consist of USRP N210 motherboards with XVRC2450
daughterboards, see \url{www.ettus.com}.
The output signal is connected to a ZRL-2400LN amplifier
to obtain a +15dBm output power with good linearity.  
The receivers consist of custom boards assembled by using amplifiers,
filters and mixers from mini-circuits, see \url{www.minicircuits.com}.
The receiver noise figure is around 10-11dB. 
During measurements very close to the base-stations an additional
10dB attenuator was inserted between the antennas and the
receiver boards in order to avoid saturation.
The boards were tuned
using attenuators to make the noise variance approach a
nominal value $\sigma^2_{\text{nominal}}$. The actual noise variance
varies up to one decibel from $\sigma^2_{\text{nominal}}$.
The value $\sigma^2_{\text{nominal}}$ is known by all nodes.
The analog output signal from the receiver boards are digitized
at an intermediate frequency of 70MHz 
by USRP N210/2 boards equipped with basic daughterboards.
A photograph of a receiver-node is shown in Fig. \ref{setup1}.
The sample-clocks of all six nodes are locked to a common 10MHz
reference and a common one pulse-per-second 
clock using long cables.
This simplifies the implementation
and gives a synchronization similar to that of a e.g. an
LTE system where all mobile-stations derive the timing
from common control channels.
However, the oscillators of the
three receiver nodes in our system are {\em not} locked to the reference
and thus there are small frequency offsets and resulting phase
rotations among the three receiver nodes and the 
transmitters. 
The base-band processing and system control is implemented
on two PCs, one for all the base-station nodes and one for all the
mobile-station nodes (the USRPs are connected with long Ethernet cables
to the PCs, each having seven Ethernet connections in all). 
The processing for each node runs in a separate thread.
The feedback from the mobile-station PC to the base-station
PC is achieved with an Ethernet cable between the two PCs.

\section{Air Interface and Signal Processing}

\begin{figure}
\begin{psfrags}
\psfrag{0}[b][b][0.5]{$0$}
\psfrag{1}[b][b][0.5]{$1$}
\psfrag{9}[b][bl][0.5]{$9$}
\psfrag{10}[b][bl][0.5]{$10$}
\psfrag{11}[b][bl][0.5]{$11$}
\psfrag{19}[b][bl][0.5]{$19$}
\psfrag{5}[b][bl][0.5]{$5$}
\psfrag{n}[b][b][0.5][90]{$n_{\text{s}}-1$}
\psfrag{p1}[b][b][0.5]{Payload data}
\psfrag{p2}[b][b][0.5]{Payload data}
\psfrag{d}[b][b][0.5]{Demodulation pilots}
\psfrag{c}[b][b][0.5]{CSI pilots}
\centerline{
   \includegraphics[width=0.4\textwidth, height=0.1\textheight ]%
     {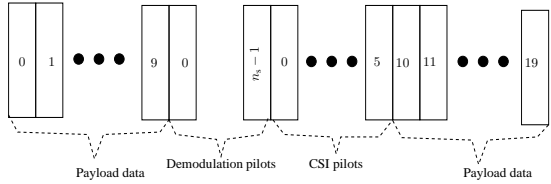}
     }
\end{psfrags}
\caption{The frame-structure}
\label{frame}
\end{figure}

The air interface is based on an OFDM modulation with 38 subcarriers,
with 312.5kHz subcarrier spacing and a cyclic prefix of 0.48$\mu$s. 
The modulation applied on each subcarrier is 16QAM. The data is encoded
in blocks of 1140 bits with a rate of 0.75 using an LDPC code.
Two coding blocks are transmitted per frame.
The frame-structure is indicated in Fig. \ref{frame}. 
During the time indicated as ``payload'', modulated symbol streams
are transmitted from all base-stations using individual
precoders (i.e., beamformers). In the section marked ``demodulation reference
pilots'', all subcarriers are occupied by known reference symbols
which have been processed by the same precoder 
as the corresponding stream.
The demodulation reference pilots are transmitted for one stream at a time
thereby avoiding any interference. 
In the area marked "CSI'', channel state information pilots are transmitted.
This means that a pilot symbol is transmitted from each of the six
antennas in the system sequentially without interference.
The mobile-stations estimate their channels independently for each
subcarrier and feed back the impulse responses to one of the
base-stations which then has ``global'' channel state information.
This (master) base-station calculates the beamformers and informs the other
base-stations of the result. The beamformers are calculated
according to the ``max-SINR'' approach described in \cite{GOM:08}.
The total power of all streams is normalized to +15dBm.
This approach is followed both in the IA and CoMP cases.
The difference being that all streams emanate from a single
six-antenna base-station in the CoMP case and three distinct
two-antenna base-stations in the IA case.

The master base-station calculates a receiving vector (combiner) for
each mobile-station. This vector is not used by the receiver.
Instead the receiver uses an MMSE vector calculated based on the 
demodulation pilot symbols of the desired and interfering streams 
and the nominal noise power.

The overhead of the CSI pilots is substantial. However, the
frame is only 0.1ms long. 
The payload of the frame could
be made much longer without the need for additional pilots
(assuming moderate mobility) - and thereby 
reduce the overhead in relative terms.
For this reason we will ignore the overheads when calculating throughput.

In the reference cases SIMO and MIMO, - no closed-form beamforming is used.
In fact, no beamforming is used at all. The SIMO and MIMO cases
exist in two variants ``TDMA'' and ``All-ON''. In the TDMA
case only one base-station is transmitting at a time
while all base-stations are active all the time in the All-ON mode.
Thus the total number of streams in the system is one in the TDMA-SIMO case,
two in the TDMA-MIMO case, three in the All-SIMO case and six
in the All-MIMO case. In the IA and CoMP cases, there are three
streams in the system. This is the maximum number of streams for IA - 
while CoMP could utilize more and thereby potentially improve performance.

\section{Measurement Campaign}

The measurements were made in 116 batches.
In each batch, all the schemes were run sequentially with
one second delay between the schemes. Each scheme was run
with five frames inter-spaced 0.1seconds.
The statistics from the first of these five frames is not used since
the base-stations have not yet received any feedback information
from the mobile-stations. The personnel involved in the measurements
were standing still during the batches in order
not to outdate the channel
state information at the transmitter and give all schemes as
similar channels as possible.

\begin{figure}
\begin{psfrags}
\psfrag{ta}[l][b][0.5]{Transmitter antennas of BS \#1}
\psfrag{ms1}[l][b][0.5]{Mobile-station \#1}
\psfrag{ms2}[b][b][0.5]{Mobile-station \#2}
\psfrag{bs2}[b][b][0.5]{Base-station \#2}
\psfrag{bs1}[b][b][0.5]{Base-station \#1}
\centerline{
   \includegraphics[width=0.4\textwidth, height=0.17\textheight ]%
     {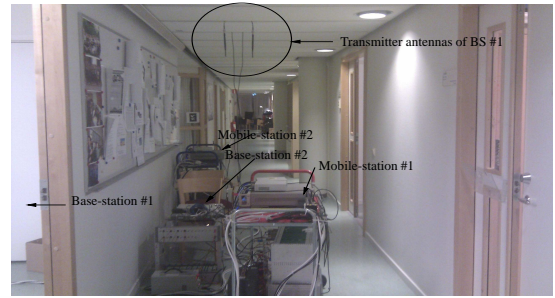}}
\end{psfrags}
\caption{The measurement environment and some equipment}
\label{setup2}
\end{figure}

\begin{figure}
\begin{psfrags}
\centerline{
   \includegraphics[width=0.35\textwidth, height=0.35\textheight ]%
     {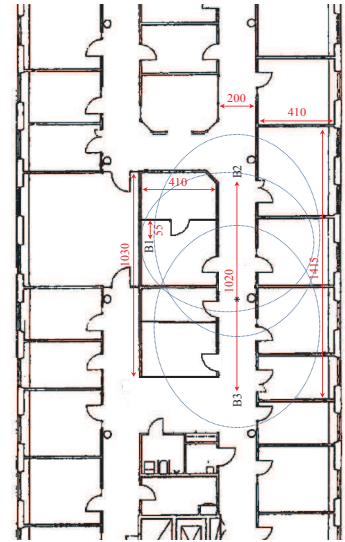}}
\end{psfrags}
\caption{Map of the measurement environment. The positions of the base-stations
are marked BS1, BS2 and BS3. The three circles indicate the areas within
which MS1, MS2 and MS3 roamed around during the measurements.}
\label{map}
\end{figure}

The measurement environment can be classified as indoor office,
see Fig. \ref{setup2}.
The three base-stations were distributed as shown in the floor-map
of Fig. \ref{map}.  The power
of the base-stations were 15dBm . The three mobile-stations were 
fixed during the batches but moved moved 
between the batches. The mobile-stations were
mostly located within the circle sorrounding it's associated
base-station in Fig. \ref{map}.

\section{Results}

In this section we present the measurement results from the
described in the previous section. We first note that the
raw signal to thermal noise ratio (i.e. without any spatial processing
and  averaged over all subcarriers)
was in the range of 32dB to 61dB. Thus the thermal noise
is almost negligable.

\subsection{Raw Results}

The transmitted data is generated from a random number generator -
the seed of which is calculated from the start time of the batch.
The mobile-stations are thus able to calculate the bit and
frame-error rates (FER). The frame
error rate is listed for all the implemented schemes
in Table \ref{FERtable}.
The reader may note that CoMP is performing significantly better
than IA in the part of the table  under ``All data''.
In this part of the data the mobile-station may
not be connected to the strongest base-station.
Thus a substantial performance improvement
can be achieved by simply handing over the mobile-station to the
strongest base-station. In order to clean the results from this
effect, the results in Table \ref{FERtable} are divided into two parts:
one where all batches are used and one where only data for mobile-stations
connected to the strongest base-station are used.%
\footnote{ For CoMP we use the same selection of measurements as for
the other schemes in order to make the results directly comparable.
}%
All the remaining results will consider the ``Best-BS'' case.

In Table \ref{FERtable} we have also plotted the  throughput
defined as $n_{\text{s}}(1-\text{FER})$ where $n_{\text{s}}$
is the number of streams in the system. 

\begin{table} 
\begin{center}
\scalebox{0.87}{
\begin{tabular}{|c|c|c|c|c|c|c|}\hline
   & \multicolumn{2}{|c|}{All data} &
  \multicolumn{4}{|c|}{Best BS} \\ \hline
Method & FER & c-FER & FER & c-FER & rate & c-rate \\ \hline
IA & 0.31 & 0.04 & 0.21 & 0.02 & 2.36 & 2.95  \\ \hline 
CoMP & 0.01 & 0.00 & 0.06 & 0.01 & 2.81 & 2.97  \\ \hline 
TDMA-MIMO & 0.08 & 0.01 & 0.04 & 0.00 & 1.93 & 2.00  \\ \hline 
TDMA-SIMO & 0.00 & 0.00 & 0.00 & 0.00 & 1.00 & 1.00  \\ \hline 
All-MIMO & 0.99 & 0.92 & 0.98 & 0.87 & 0.13 & 0.78  \\ \hline 
All-SIMO & 0.76 & 0.55 & 0.61 & 0.31 & 1.18 & 2.07  \\ \hline
\end{tabular}
}
\end{center}
\caption{FER: Raw frame error rate, c-FER: frame error rate
of coded bits, rate: average number of correct frames times
the number of streams, c-rate: as rate but with coded bits,
All-data: all batches are used, Best-BS only batches where the
MS is connected to the strongest BS.}
\label{FERtable}
\end{table}

From Table \ref{FERtable} we conclude that the performance
of CoMP is better than IA for uncoded transmissions.
For coded transmissions both IA and CoMP exhibit a
FER very near zero.
The reference schemes SIMO and MIMO are all worse than IA and CoMP, -
and reach only 70\% of their rate.

\subsection{Comparison: Measurements against Theory}

An important aspect of experimentation is to verify the models
used for system simulations. This is important since we are unable
to experimentally investigate every relevant scenario - of propagation
environment, user distribution, traffic loads, algorithm parameters
and so on. Therefore we focus on quantifying the difference between
the performance we would have predicted for the scenario at hand
and the performance we actually obtained. The result of this
analysis in illustrated in Fig. \ref{SINDR_IA},
\ref{SINDR_COMP} and \ref{SINDR_SIMO} for the
IA, COMP and TDMA-SIMO cases, respectively.
The details of this analysis is described below.

\subsubsection{EVM and Performance Modelling}

\begin{figure}
\begin{psfrags}
\centerline{
\scalebox{0.4}{\includegraphics*[1.5in,1.5in][8.7in,6.5in]%
     {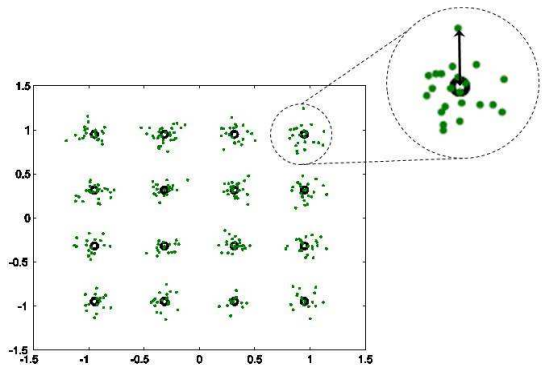}}}
\caption{Illustration of error vector}
\label{CONST_IA}
\end{psfrags}
\end{figure}

We start-off with the performance that may be
predicted given the measurements from the CSI pilots only. 
Based on these measurements we calculate the beamforming vectors
of our beamforming strategy according to \cite{GOM:08}
and obtain ``post-processed'' signal to interference and noise
ratios (SINR-post) at the output of the receiver combining 
(i.e., the quality of the equvivalent SISO channels formed 
by transmit beamformer and receive combiners).%
\footnote{Every bit from each A/D converter collected during the
measurement is stored and made available for post-processing.
We are therefore able to perform the post-processing described in this
section.}%
The SINR-post factor is finally obtained as an average over the subcarriers
calculated as

\begin{equation}
\text{SINR-post} = \frac{\sum_i S_i}{\sum_i I_i + \sigma^2_{\text{nominal}}},
\label{eq_av}
\end{equation}

where $S_i$,$I_i$ and $\sigma^2_{\text{nominal}}$ are the signal, interference
and nominal noise power on the $i$th subcarrier in a certain frame
(the noise power is assumed identical on all subcarriers).
The CDF of the resulting SINR-post is plotted in Fig. 
\ref{SINDR_IA}-\ref{SINDR_SIMO} and is marked with the legend ``ideal''.
The above calculation neglected the channel estimation errors
and the fact that the channel may change
between frames. In other words, it is non-causal as the channel state
information used in the calculation of the beamformers is actually not
available until after the time the frame has been transmitted.
As a next step we therefore replace the channel state used
in the transmit beamformers with channel state available
in the previous frame. The result is also shown in 
Fig. \ref{SINDR_IA}-\ref{SINDR_SIMO} and marked with ``causal''.
Is this the real quality of the channel as seen by the 
mobile-station? - no it's not.
The real quality of the channel seen from the view-point
of the SISO modem (which is transmitting over the 
equivalent SISO channel created by transmitter precoding 
and receiver combining) 
is best represented by the error vector magnitude.
The error vector is defined as the difference between
the receive constellation points and the true constellation
points as illustrated in Fig. \ref{CONST_IA}.
The error vector magnitude is defined as root of the variance of the
error vector, normalized by the power of the constellation
positions. To compare this value with the 
previously calculated
SINR values  we form the following EVM based SINDR
estimate

\begin{equation}
\text{SINDR}_{\text{EVM}} = \sum_{i} p_i \text{EVM}_i^{-2},
\label{SINRevm}
\end{equation}

where $p_i$ is the power of the virtual SISO channel on subcarrier
$i$ and $\text{EVM}_i$ is the EVM of the corresponding subcarrier.
We use this power-weighted EVM value as it corresponds better to
the average SINR value defined in (\ref{eq_av}) than a straight
average. Note that we have used the acronym SINDR in 
(\ref{SINRevm}). This acronym denotes ``signal to noise, interference
and distortion ratio'' in order to emphasize that the 
$\text{SINDR}_{\text{EVM}}$ measure will include also the dirty-RF
impairments caused by phase-noise and non-linearities i.e.
the ``distortions''.

The curves labelled ``EVM-model'' in Fig. \ref{SINDR_IA}-\ref{SINDR_SIMO} 
are obtained 
by using the same non-causal channel matrices as ``ideal''.
However, this model includes also the error model used
in \cite{ZET:10a} and a common phase error.
In \cite{ZET:10a} a Gaussian noise term is added
to each transmitter and receiver antenna branch.
The power of this modeled noise is set to 34dB below the desired
signal in the transmitter and 40dB below in the receiver.
These values have been estimated from the SISO
measurements and from the data-sheet of the MAX2829
circuit, \url{http://www.maxim-ic.com/datasheet/index.mvp/id/4532}, 
used in the XCVR2450 daughterboard.
Note that this error will affect both the payload symbols
and the training. 
In addition to the model of \cite{ZET:10a} we also introduce
a so-called common phase rotation, see \cite{COR:98}.
This is a phase-error which (despite it's name) will
be independent between all our six transmitter branches 
(since each has its own local oscillator). 
This phase error will be introduced as a random rotation
of the phase of the six transmitter branches which is
re-randomized between frames. The phase is set to Gaussian with standard
deviation of 0.6 degrees, again based on the data-sheet of the
MAX2820 circuit.

\subsubsection{Discussion}

\begin{figure}
\begin{psfrags}
\psfrag{a}[lt][r][0.5]{$\text{SINR-post}$ ideal}
\psfrag{b}[lt][r][0.5]{$\text{SINR-post}$ causal}
\psfrag{c}[lt][r][0.5]{$\text{SINDR}_{\text{EVM}}$ measured}
\psfrag{d}[lt][r][0.5]{$\text{SINDR}_{\text{EVM}}$ model}
\centerline{
   \includegraphics[width=0.4\textwidth, height=0.25\textheight ]%
     {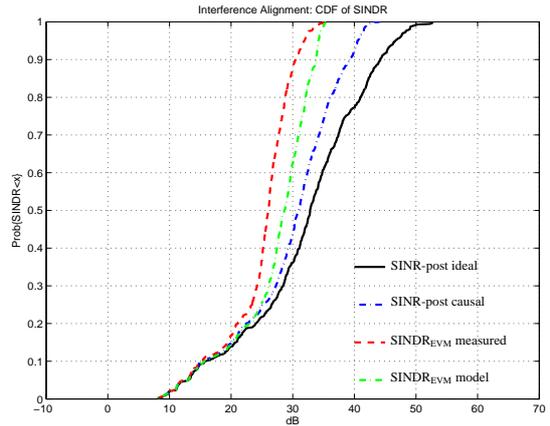}}
\caption{Interference-alignment: Distribution of signal to interference, noise, and distortion
ratio (SINDR) based on different models and measurements}
\label{SINDR_IA}
\end{psfrags}
\end{figure}

\begin{figure}
\begin{psfrags}
\psfrag{a}[lt][r][0.5]{$\text{SINR-post}$ ideal}
\psfrag{b}[lt][r][0.5]{$\text{SINR-post}$ causal}
\psfrag{c}[lt][r][0.5]{$\text{SINDR}_{\text{EVM}}$ measured}
\psfrag{d}[lt][r][0.5]{$\text{SINDR}_{\text{EVM}}$ model}
\centerline{
   \includegraphics[width=0.4\textwidth, height=0.25\textheight ]%
     {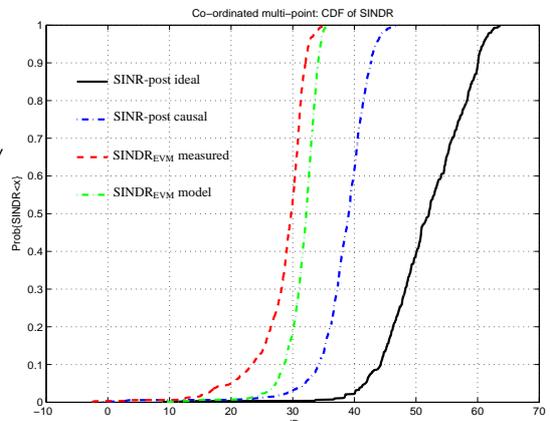}}
\caption{Coordinated Multi-Point: Distribution of signal to interference, noise, and distortion ratio (SINDR) based on different models and measurements}
\label{SINDR_COMP}
\end{psfrags}
\end{figure}

\begin{figure}
\begin{psfrags}
\psfrag{a}[lt][r][0.5]{$\text{SINR-post}$ ideal}
\psfrag{b}[lt][r][0.5]{$\text{SINR-post}$ causal}
\psfrag{c}[lt][r][0.5]{$\text{SINDR}_{\text{EVM}}$ measured}
\psfrag{d}[lt][r][0.5]{$\text{SINDR}_{\text{EVM}}$ model}
\centerline{
   \includegraphics[width=0.4\textwidth, height=0.25\textheight ]%
     {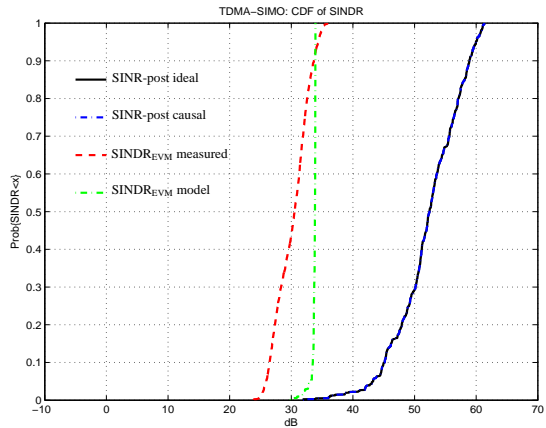}}
\caption{TDMA SIMO: Distribution of signal to interference, noise, and distortion ratio (SINDR) based on different models and measurements}
\label{SINDR_SIMO}
\end{psfrags}
\end{figure}

From Fig. \ref{SINDR_IA}-\ref{SINDR_SIMO} we conclude that
our EVM-model is able to bridge most of the gap between the
``ideal'' and ``EVM model'' results. This indicates that most of the
degradation between the ``ideal'' and ``causal'' curves
are not due to the propagation channel evolution during the
consecutive time-slots but due to dirty-RF effects. 
However, there is still some difference between the
``EVM model'' and the actual EVM measurements.
Part of this difference could be due to channel evolution,
but not all of it. Since there is also a mismatch
in the SIMO case. 
A detailed studied of the actual performance of the hardware
could help reducing the gap further.

Finally, we ask the reader to notice the extremely high
performance of the CoMP scheme in the ``ideal'' case
(far superior over the IA ``ideal'' case) .
However, this performance advantage diminishes into 
a more modest advantage in reality.
This implies that CoMP is very susceptible to any
kind of non-idealness.
Sensitivity of CoMP to common phase rotation was also recently
studied in \cite{Bjornson2011a}.

\section{Conclusions}

We have implemented interference alignment (IA) and coordinated multi-point
(CoMP) on a wireless testbed. We observe an performance improvement
over reference schemes such as SIMO and MIMO. However, the gains
are much smaller than what could be theoretically calculated
on the basis from our channel estimates. The reason being that
dirty-RF effects come into play and substantially degrade
the performance. We are able to model the dirty-RF effects reasonably
well with simple models but there is still room for improvement.
The performance of CoMP is the highest of the implemented schemes.
However, the performance advantage is not as great as predicted
by theory. Both CoMP and IA perform better than the reference
schemes of single-user SIMO and MIMO.



\end{document}